\documentclass[%
reprint,
amsmath,amssymb,
aps,
]{revtex4-1}

\usepackage{graphicx}
\usepackage{dcolumn}
\usepackage{bm}
\usepackage[caption=false]{subfig}

\begin{document}
	
	\preprint{INTEL-CONFIDENTIAL}
	
	\title{Physical Origin of Transient Negative Capacitance in a Ferroelectric Capacitor}
	
	\author{Sou-Chi Chang}
	\email{sou-chi.chang@intel.com}
	\affiliation{Components Research, Intel Corporation, Hillsboro, Oregon 97124, USA}
	
	\author{Uygar E. Avci}%
	\affiliation{Components Research, Intel Corporation, Hillsboro, Oregon 97124, USA}
	
	\author{Dmitri E. Nikonov}%
	\affiliation{Components Research, Intel Corporation, Hillsboro, Oregon 97124, USA}
	
	\author{Sasikanth Manipatruni}%
	\affiliation{Components Research, Intel Corporation, Hillsboro, Oregon 97124, USA}
	
	\author{Ian A. Young}%
	\affiliation{Components Research, Intel Corporation, Hillsboro, Oregon 97124, USA}

	\begin{abstract}
	Transient negative differential capacitance (NC), the dynamic reversal of transient capacitance in an electrical circuit is of highly technological and scientific interest since it probes the foundation of ferroelectricity. In this letter, we study a resistor-ferroelectric capacitor (R-FeC) network through a series of coupled equations based on Kirchhoff’s law, Electrostatics, and Landau theory. We show that transient NC in a R-FeC circuit originates from the mismatch between rate of free charge change on the metal plate and that of bound charge change in a ferroelectric (FE) capacitor during polarization switching. This transient charge dynamic mismatch is driven by the negative curvature of the FE free energy landscape. It is also analytically shown that a free energy profile with the negative curvature is the only physical system that can describe transient NC during the two-state switching in a FE capacitor. Furthermore, this transient charge dynamic mismatch is justified by the dependence of external resistance and intrinsic FE viscosity coefficient. The depolarization effect on FE capacitors also shows the importance of negative curvature to transient NC. The relation between transient NC and negative curvature provides a direct insight into the free energy landscape during the FE switching.
	\end{abstract}
	
	
	\maketitle
	
	
	In the past four decades, the computing power of microprocessors has been significantly improved due to the relentless pursuit of Moore's law \cite{658762}. However, the static power becomes increasingly important in the total energy dissipation as the complementary metal-oxide-semiconductor (CMOS) transistors in a microchip are scaled down to the nanometer regime due to the reduction of on-off current ratio \cite{1250885}. Recently, a novel gate structure based on the FE oxide (also known as a negative capacitance field-effect-transistor) has been proposed for transistors to improve the on-off current ratio by enhancing depolarization fields in the gate stack  \cite{doi:10.1021/nl071804g,8027211}. The main idea behind this approach is the elimination of negative curvature in the FE thermodynamic profile by depolarization, which can be achieved by engineering the dielectric and FE thicknesses in the gate stack. Therefore, it is of importance to show the existence of negative curvature between two polarization states in the thermodynamic free energy profile as predicted by Landau theory experimentally.
	
	Recently, it has been claimed that the transient NC measured in a R-FeC circuit can be viewed as a direct mapping to the negative curvature of FE free energy profile during polarization switching \cite{Khan2015,ADFM:ADFM201602869}. However, so far there is no clear physical and theoretical description to directly link transient NC to the negative curvature of FE free energy profile. Hence, it is of importance to establish a correct physical picture for the transient NC measured in a R-FeC circuit \cite{Khan2015,ADFM:ADFM201602869} and also its relation to the negative curvature of FE free energy landscape. In this letter, we show that transient NC measured in a R-FeC circuit originates from the mismatch between rate of free charge change on the metal plate and that of bound charge change in a FE capacitor during polarization switching, which is driven by the negative curvature of thermodynamic profile of the FE. Analytical expressions are also provided to show that the negative curvature in the free energy profile is the only solution that can physically describe transient NC during the two-state switching in a R-FeC circuit. 
	
	\begin{figure}
		\begin{center}
				\includegraphics[width=0.6\linewidth]{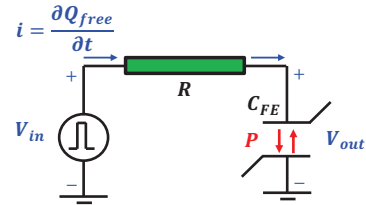}%
			\caption{The schematic of a resistance-ferroelectric capacitor (R-FeC) circuit for studying the physical origin of transient negative capacitance (NC) observed in the experiments in Refs. \cite{Khan2015,ADFM:ADFM201602869}. $V_{in}$, $V_{out}$ are input and output voltage, respectively. $R$ is external resistance. $C_{FE}$ is the ferroelectric capacitor. The current flowing through the resistor is determined by the first-order time derivative of free charge, $Q_{free}$.}
			\label{fig1}
		\end{center}
	\end{figure}
	
	\begin{figure*}
		\begin{center}
			\subfloat[]{%
				\includegraphics[width=.4\linewidth]{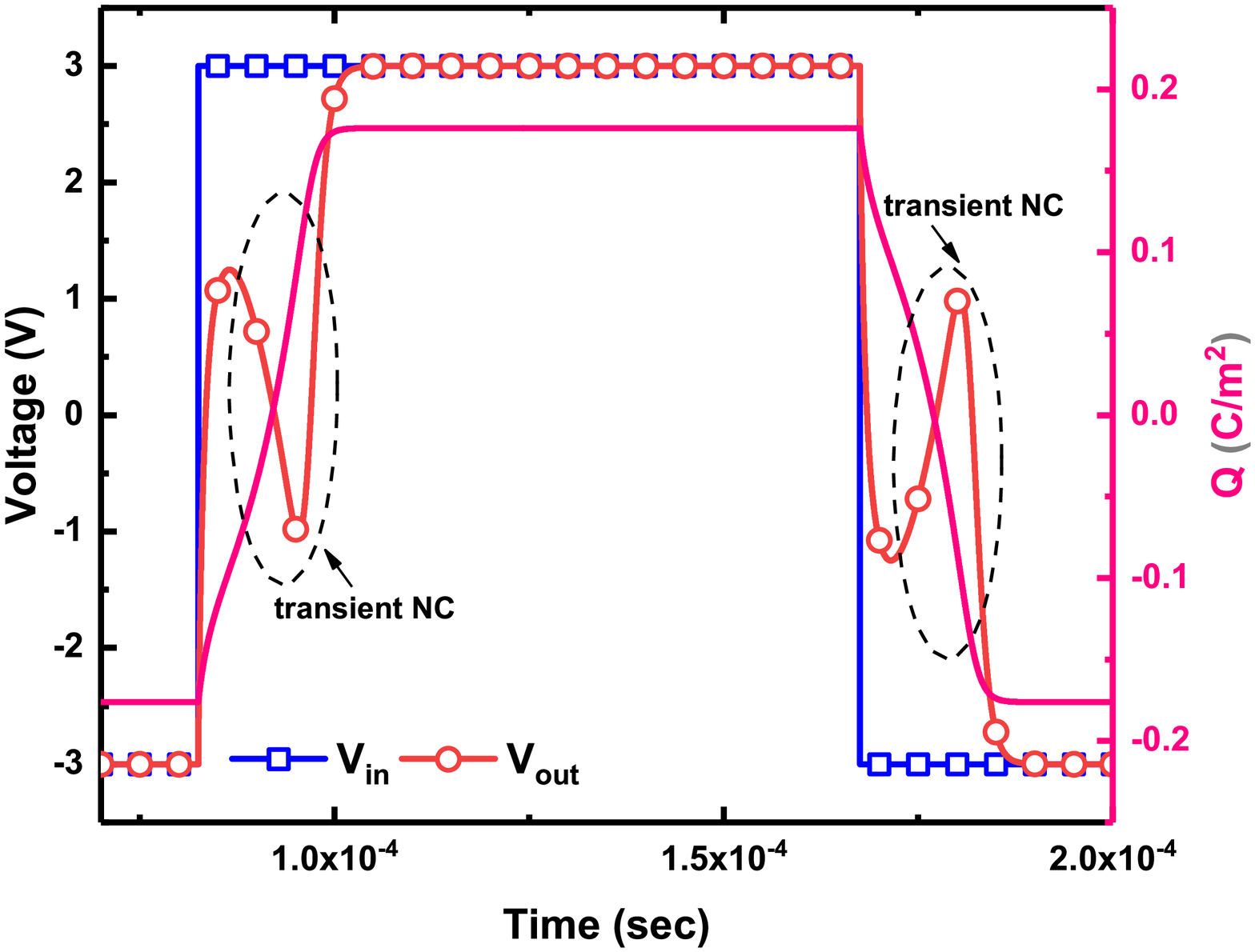}%
			}
			\subfloat[]{%
				\includegraphics[width=.4\linewidth]{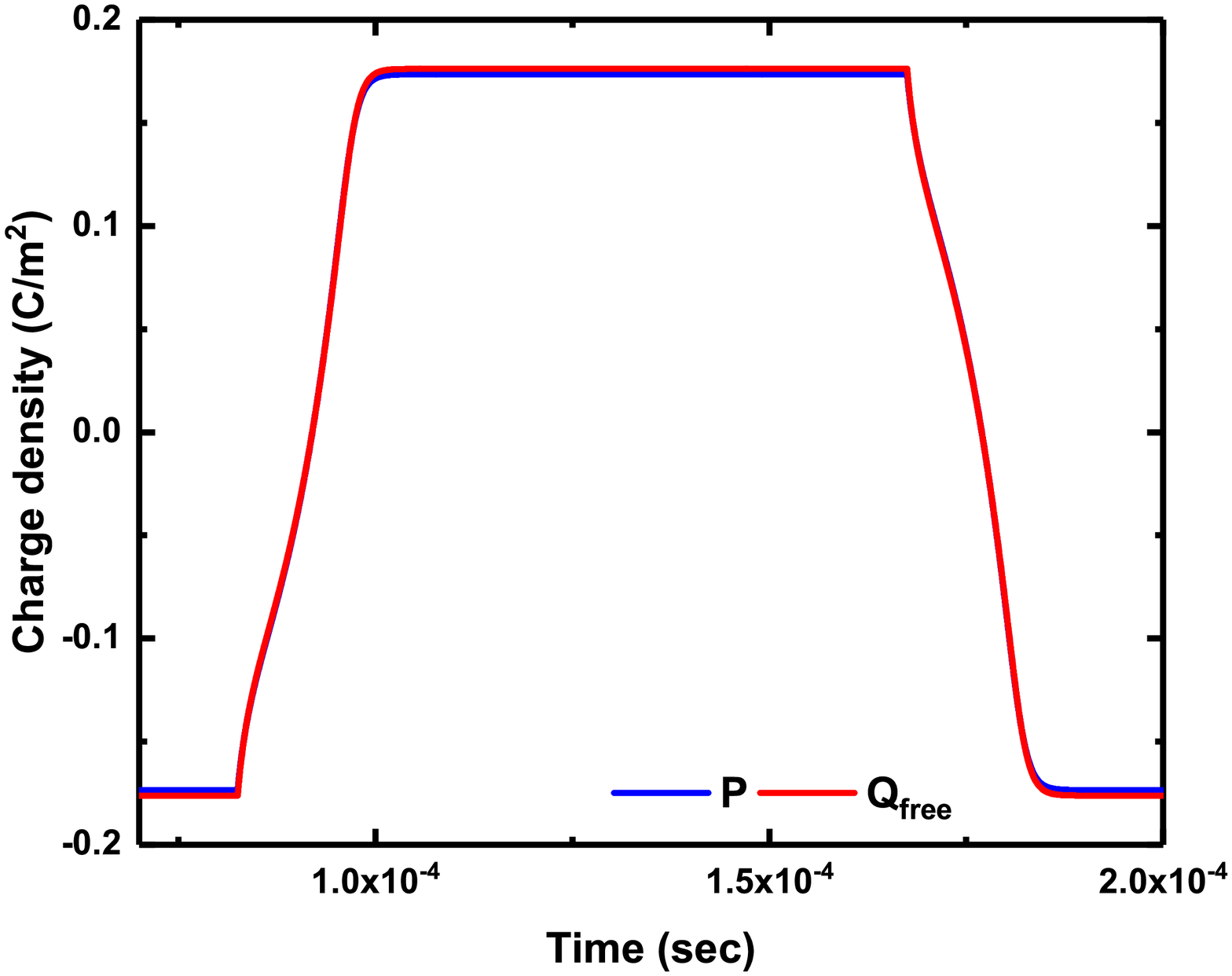}%
			} \\
			\subfloat[]{%
				\includegraphics[width=.4\linewidth]{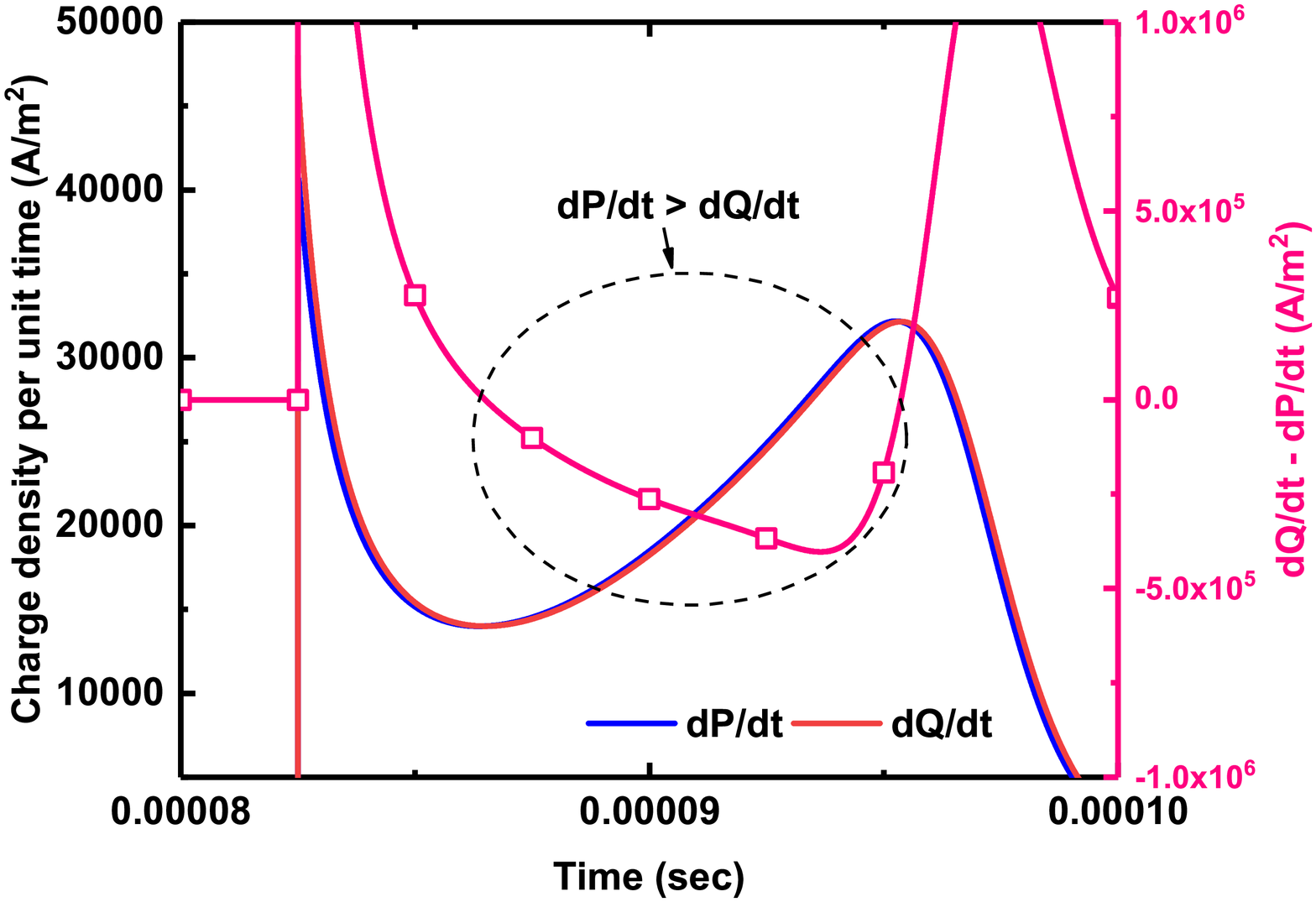}%
			} 
			\subfloat[]{%
				\includegraphics[width=.4\linewidth]{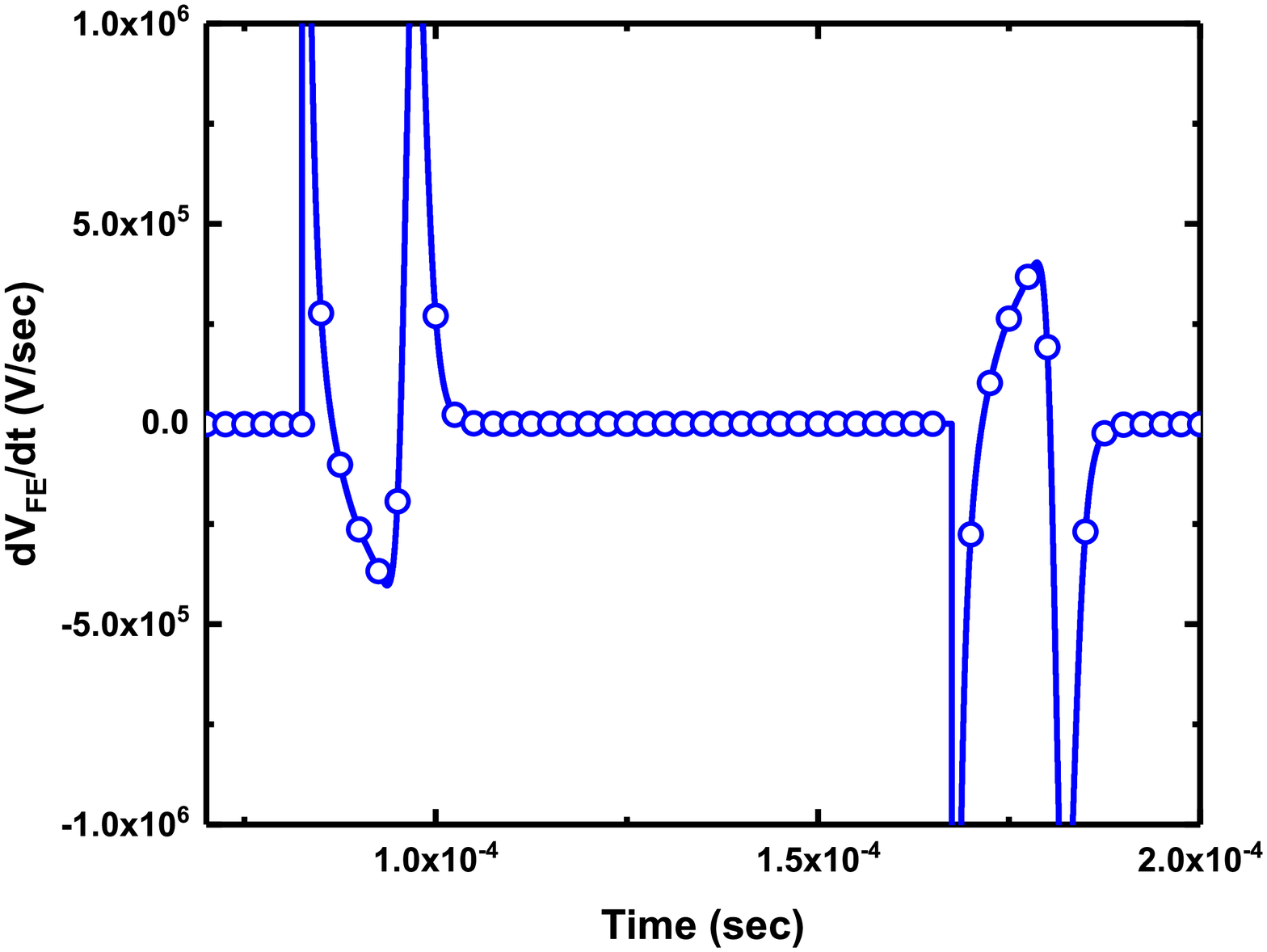}%
			}
			\caption{(a) Input voltage (blue), output voltage (red), and free charge on a FE capacitor (pink) as functions of time. (b) Polarization and free charge as functions of time. (c) Charge density per unit time for free charge (red) and polarization (blue). The difference between $\frac{\partial Q}{\partial t}$ and $\frac{\partial P}{\partial t}$ is given in the pink. (d) Change in the voltage across a FE capacitor per unit time as a function of time.}
			\label{fig2}
		\end{center}
	\end{figure*}
	
	To describe the dynamics for both free charge and polarization in a R-FeC circuit as shown in Fig. \ref{fig1}, (i) Kirchhoff's law is used to describe the displacement current flowing through an external resistor, (ii) Electrostatics is applied for the fact that net charge on a FE capacitor has to be equal to free charge plus bound charge, and (iii) polarization dynamics under an electric field across the FE oxide is captured by Landau theory under the mono-domain approximation. The equation corresponding to Kirchhoff's law is given as
	\begin{eqnarray}
	\frac{\partial Q_{free}}{\partial t} &=& \frac{V_{in}-V_{out}}{RA}=\frac{V_{in}-V_{FE}}{RA},
	\label{eq1}
	\end{eqnarray}
	where $\frac{\partial Q_{free}}{\partial t}$ is the displacement current density (A/m$^{2}$), $V_{in}$ and $V_{out}$ are the input and output voltages ($V$), respectively, $R$ is the external resistance ($\Omega$), $A$ is the area of a capacitor (m$^{2}$). Note that $V_{out}$ is equal to $V_{FE}$, which is the voltage across a FE capacitor. Based on the Electrostatics, the free charge density ($Q_{free}$) on a FE capacitor can be written as
	\begin{eqnarray}
	Q_{free}&=&\epsilon_{0}E_{FE} + P,
	\label{eq2}
	\end{eqnarray}
	in which $E_{FE}$ is the electric field across the FE oxide, $\epsilon_{0}$ is the vacuum dielectric constant, and $P$ is the FE polarization, whose single-domain dynamics is governed by the Landau theory given as \cite{PhysRevB.20.1065,7373582,PhysRevApplied.4.044014,PhysRevB.88.024106,PhysRevB.68.094113}
	\begin{eqnarray}
	\gamma\frac{\partial P}{\partial t}&=&-\left(2\alpha_{1}+\frac{t_{DEP}}{\epsilon_{0}t_{FE}}\right)P-4\alpha_{11}P^{3}-6\alpha_{111}P^{5} \nonumber \\ 
	& & + E_{FE},
	\label{eq3}
	\end{eqnarray}
	where $\alpha_{1}$, $\alpha_{11}$, and $\alpha_{111}$ are thermodynamic expansion coefficients for bulk FE free energy, $\gamma$ is the viscosity coefficient, and $t_{FE}$ as well as $t_{DEP}$ are FE oxide and effective depolarization thicknesses, respectively. Note that $t_{DEP}$ is used to represent the depolarization effect associated with finite screening length of metal contacts and inevitable interface dead layers in a FE capacitor; that is, $t_{dep} = \frac{\lambda_{1}}{\epsilon_{1}}+\frac{\lambda_{2}}{\epsilon_{2}}+\frac{t_{DE}}{\epsilon_{DE}}$ with $\lambda_{1}$, $\lambda_{2}$, and $t_{DE}$ being the screening length of top and bottom metal contacts and dead layer thickness, respectively, and $\epsilon_{1}$, $\epsilon_{2}$, and $\epsilon_{DE}$ are the relative dielectric constants of top and bottom contacts and dead layer, respectively \cite{PhysRevApplied.7.024005}. By combining Eqs. \ref{eq1} to \ref{eq3}, the differential equation for free charge dynamics in a R-FeC circuit is given as
	\begin{eqnarray}
	RA\frac{\partial Q_{free}}{\partial t} &=& V_{in} - \frac{t_{FE}\left(Q_{free}-P\right)}{\epsilon_{0}}, \label{eq4}
	\end{eqnarray}
	From Eqs. \ref{eq3} and \ref{eq4}, one can immediately see that free charge and polarization are coupled through the electric field across the FE oxide. Also, the free charge and polarization responses are limited by the external resistance and intrinsic viscosity coefficient, respectively.
	\begin{table}
		\caption{\label{tab1} Simulation parameters for a R-FeC circuit.}
		\begin{tabular}{c c c}
			\hline
			\hline
			Symbol & Quantity & Value \\
			\hline
			$\alpha_{1}$ &  & $-1.05\times10^{9}\frac{m}{F}$ \cite{7458805} \\
			$\alpha_{11}$ & Landau expansion coefficient & $10^{7} \frac{m^{5}}{C^{2}F}$ \cite{7458805} \\
			$\alpha_{111}$ & & $6\times10^{11}\frac{m^{9}}{C^{4}F}$ \cite{7458805} \\
			$t_{dep}$ & Effective depolarization thickness & $0.05$nm \\
			$t_{FE}$ & Ferroelectric thickness & $10$nm \\
			$A$ & Capacitor area & $50^{2}\mu$m$^{2}$ \\
			$R$ & External resistance & 50k$\Omega$ \\
			$\gamma$ & Viscosity coefficient & $5\times10^{2}$ m$\cdot$sec/F \\
			\hline
			\hline	
		\end{tabular}
	\end{table}
	
	The simulation parameters are summarized in Table \ref{tab1} if not mentioned elsewhere. Eqs. \ref{eq3} and \ref{eq4} are integrated numerically using the Euler method, and the time step is chosen such that the results remain unchanged with a shorter time step. Note that for all the simulations shown below, initially a negative input voltage is applied on the capacitor to make the polarization in the negative direction.
	\begin{figure}
		\begin{center}
			\subfloat[]{%
				\includegraphics[width=0.9\linewidth]{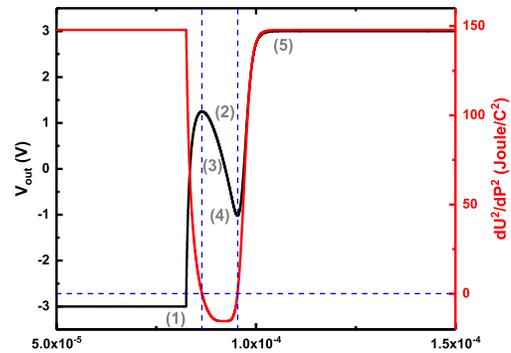}%
			} \\
			\subfloat[]{%
				\includegraphics[width=0.9\linewidth]{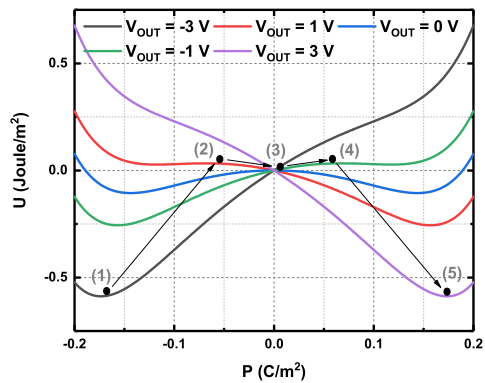}%
			} 
			\caption{(a) Output voltage and curvature of FE thermodynamic free energy profile as functions of time. The sign reversal of $\frac{\partial V_{out}}{\partial t}$ coincides with $\frac{\partial^{2} U}{\partial P^{2}}<0$ (b) The thermodynamic free energy profiles at different time steps for charging a FE capacitor. Black dots and arrows connecting the black dots represent transient polarization states and switching directions, respectively. The parameters are the same as those in Fig. \ref{fig2}.}
			\label{fig5}
		\end{center}
	\end{figure}
	
	Fig. \ref{fig2}(a) shows the output response in a R-FeC circuit under an input pulsed signal. The free charge on a FE capacitor is also given in the same figure. From Fig. \ref{fig2}(a), it can be seen that there are two regions where the free charge is increased (decreased) but the voltage is decreased (increased); that is, $\frac{\partial Q_{free}}{\partial V}< 0$. This negative differential capacitance occurs only during the polarization reversal, which can be seen at the same time scale in Fig. \ref{fig2}(b), where both free charge and polarization as functions of time are given. Fig. \ref{fig2}(b) also indicates that the free charge and polarization are almost equal in a R-FeC circuit under a pulsed input. However, Fig. \ref{fig2}(c) shows that there always exists a small difference between free charge and polarization in the ramping rate while charing a FE capacitor, and $\frac{\partial P}{\partial t}>\frac{\partial Q_{free}}{\partial t}$ occurs only when the polarization is switched. From Eq. \ref{eq6}, it can be seen that how the voltage across a FE capacitor changes with time is linearly proportional to the mismatch between free charge and polarization. By comparing Figs. \ref{fig2}(a) and (d), it can be seen that transient NC shows up exactly when $\frac{\partial V_{FE}}{\partial t}$ becomes negative, which can be physically explained by the fact that the polarization switching is too fast such that the free charge limited by a R-FeC circuit cannot follow perfectly \cite{Catalan2015,8027211}. 
	\begin{eqnarray}
	\frac{\partial V_{FE}}{\partial t} = \frac{t_{FE}}{\epsilon_{0}}\left(\frac{\partial Q_{free}}{\partial t}-\frac{\partial P}{\partial t}\right)
	\label{eq6}
	\end{eqnarray}

		\begin{figure}
		\begin{center}
			\subfloat[]{%
				\includegraphics[width=0.9\linewidth]{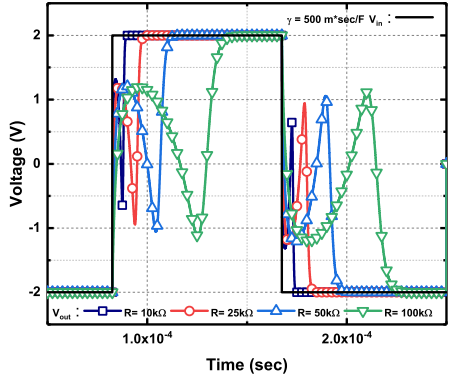}%
			} \\
			\subfloat[]{%
				\includegraphics[width=0.9\linewidth]{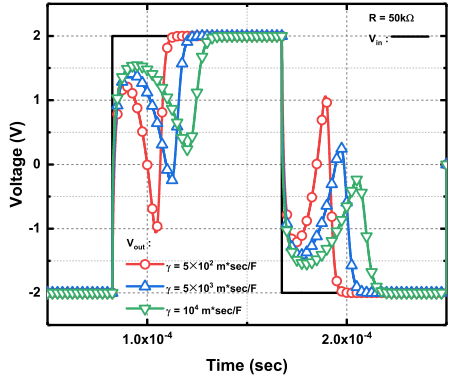}%
			} 
			\caption{The effect of (a) external resistance and (b) viscosity coefficient on transient negative capacitance in a R-FeC circuit.}
			\label{fig3}
		\end{center}
		\end{figure}
	
		\begin{figure}
			\begin{center}
			\subfloat[]{%
				\includegraphics[width=0.9\linewidth]{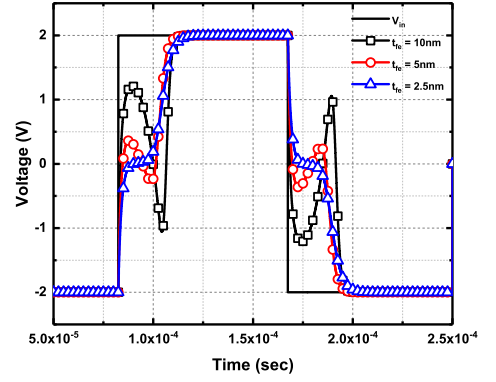}%
			} \\
			\subfloat[]{%
				\includegraphics[width=0.9\linewidth]{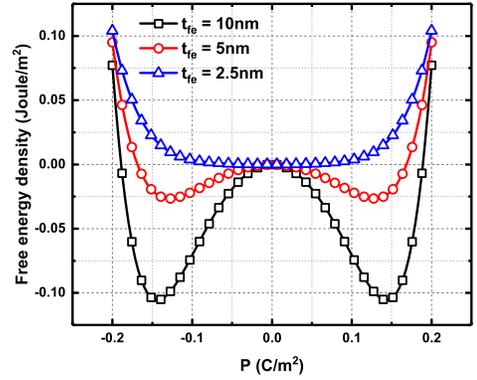}%
			}
		\caption{(a) Input and output voltages as functions of time with different FE thicknesses ($t_{dep}=0.05$nm). (b) Thermodynamic free energy profile of FE with different FE thicknesses ($t_{dep}=0.05$nm).}
			\label{fig4}
		\end{center}
	\end{figure}

	Next, the relation between free charge-polarization mismatch and negative curvature of thermodynamic profile is discussed. Fig. \ref{fig5}(a) shows that transient NC occurs when the polarization is in the negative curvature region of thermodynamic energy profile. This correlation can be also seen in Fig. \ref{fig5}(b), where the thermodynamic profiles at different time steps are given while charging a FE capacitor. From Fig. \ref{fig5}(b), it is found that the FE free energy profile is always changing as the free charge is built up across the capacitor. Therefore, Fig. 1 in Ref. \cite{Khan2015} is not a rigorous physical picture to describe transient NC in a R-FeC circuit. Note that in Fig. \ref{fig5}(a), the corresponding negative curvature in the region of transient NC can be considered as a direct mapping of negative curvature in the free energy profile under equilibrium, since from Eq. \ref{eq3}, $\frac{\partial^{2}P}{\partial U^{2}}$ has no dependence on the electric field across the FE capacitor. Furthermore, from Eq. \ref{eq6}, it can be seen directly that when a capacitor is charged with a positive input voltage, transient NC appears when $\frac{\partial P}{\partial t}>\frac{\partial Q_{free}}{\partial t}>0$ is satisfied. On the other hand, $\frac{\partial P}{\partial t}<\frac{\partial Q_{free}}{\partial t}<0$ is required for transient NC as a negative input voltage is applied in a R-FeC circuit. 
	
	Based on Landau theory ($\gamma\frac{\partial P}{\partial t}=-\frac{\partial U}{\partial P}$ with $U$ being the free energy), $\frac{\partial V_{FE}}{\partial t}$ at a given time, $t_{0}$, can be connected to the curvature of thermodynamic energy profile as shown in Eq. \ref{eq7}.
	\begin{eqnarray}
	\frac{\partial V_{FE}}{\partial t}\mid_{t=t_{0}} &=& \frac{t_{FE}}{\epsilon_{0}}\left(\frac{\partial Q_{free}}{\partial t}\mid_{t=t_{0}}+\frac{\partial U}{\gamma \partial P}\mid_{t=t_{0}}\right) \nonumber \\
	&=& \frac{t_{FE}}{\epsilon_{0}}\left(\frac{\partial Q_{free}}{\partial t}\mid_{t=t_{0}}+\frac{\partial U}{\gamma \partial P}\mid_{P=P_{i}}\right. \nonumber \\
	& & \left. +\frac{1}{\gamma}\int_{P_{i}}^{P_{t=t_{0}}} \frac{\partial^{2} U}{\partial^{2} P} dP\right),
	\label{eq7}
	\end{eqnarray}
	where $P_{i}$ and $P_{t=t_{0}}$ are initial polarization and polarization at $t=t_{0}$. From Eq. \ref{eq7}, one can immediately see that there are only two possible situations (Eqs. \ref{eq8} and \ref{eq9}) for a two-state system that can satisfy the conditions for transient NC mentioned above.
	\begin{eqnarray}
	\textrm{(i)} \quad \frac{\partial^{2} U}{\partial^{2} P}\mid_{t=t_{0}}<0 \quad \textrm{for} \quad P_{i}&<&P\left(t=t_{0}\right) \quad \textrm{and} \quad V_{in}>0 \nonumber \\
	\textrm{for} \quad P_{i}&>&P\left(t=t_{0}\right) \quad \textrm{and} \quad V_{in}<0 \label{eq8} \\
	\textrm{(ii)} \quad \frac{\partial^{2} U}{\partial^{2} P}\mid_{t=t_{0}}>0 \quad \textrm{for} \quad P_{i}&>&P\left(t=t_{0}\right) \quad \textrm{and} \quad V_{in}>0 \nonumber \\
	\textrm{for} \quad P_{i}&<&P\left(t=t_{0}\right) \quad \textrm{and} \quad V_{in}<0
	\label{eq9}
	\end{eqnarray}
	From Eqs. \ref{eq8} and \ref{eq9}, it can be seen that only the condition (i) can physically describe polarization switching when charging a FE capacitor in a R-FeC circuit. As a result, the measured transient NC in a R-FeC circuit can be considered as a strong evidence showing the existence of negative curvature in the FE thermodynamic profile. Note that $\frac{\partial Q_{free}}{\partial t}$ and $\frac{\partial V_{FE}}{\partial t}$ can be directly obtained from the transient NC measurement, and thus based on Eqs. \ref{eq6} and \ref{eq7}, $P$, $\frac{\partial P}{\partial t}$, $\frac{\partial U}{\partial P}$, and $\frac{\partial^{2}U}{\partial P^{2}}$ are also known at any time step during the measurement. Consequently, the transient NC measurement in a R-FeC circuit can be served as an useful tool to directly map the Landau energy landscape of the FE.
	
	As mentioned previously, transient NC in a R-FeC circuit is due to the difference between polarization and free charge in charging rate while the polarization is reversed. It is intuitive to expect that both external resistance and intrinsic FE viscosity coefficient can play significant roles in transient NC as well. Fig. \ref{fig3}(a) illustrates the effects of external resistance on transient NC. From Fig. \ref{fig3}(a), it is found that as the external resistance becomes larger, the polarization switching is slower since it takes longer time for the free charge to build up the electric field across a FE capacitor. More importantly, transient NC becomes more pronounced with larger external resistance due to decreasing $\frac{\partial Q}{\partial t}$ (or increasing the mismatch in Eq. \ref{eq6}). On the other hand, the mismatch can also be enhanced by speeding up the polarization switching as given in Fig. \ref{fig3}(b), where a smaller viscosity coefficient refers to faster switching dynamics.
	
	So far, we have shown that the negative curvature is required to induce the mismatch between polarization and free charge (transient NC) in a R-FeC circuit. Consequently, it is intuitive to expect that the depolarization in a FE capacitor can also affect transient NC significantly. It is well known that as the FE oxide gets thinner in a capacitor, the depolarization effect becomes more pronounced to degrade ferroelectricity in the material \cite{doi:10.1063/1.4794865}. As a result, $\frac{\partial P}{\partial t}$ becomes less important in Eq. \ref{eq6} with a stronger depolarization field and thus transient NC is weaker as shown in Fig. \ref{fig4}(a). Note that transient NC can be totally removed when the depolarization field is strong enough. From Fig. \ref{fig4}(b), it can be seen that the elimination of transient NC corresponds to the transition from the two- to single-state system, which again emphasizes the importance of negative curvature in the FE thermodynamic energy profile. Note that the effect of depolarization field can significantly change the magnitude of transient NC without affecting the delay; thus, by fitting the experimental data \cite{Khan2015} via adjusting the effective depolarization thickness, transient NC in a R-FeC circuit can also be served as a platform for the depolarization field characterization. 
	
	In summary, this letter shows that transient NC in a R-FeC circuit is induced by the mismatch of charging rate between free charge and FE bound charge during polarization switching, which is driven by the negative curvature of thermodynamic profile. It is proved analytically that the negative curvature is the only solution that can physically describe the charging behavior of a FE capacitor in a R-FeC circuit. Furthermore, the dependence of resistance, viscosity coefficient, and depolarization on transient NC is shown to emphasize the importance of free charge-polarization mismatch and negative curvature of thermodynamic profile to transient NC. 
	
	\bibliography{apssamp}

\providecommand{\noopsort}[1]{}\providecommand{\singleletter}[1]{#1}%
\begin{thebibliography}{15}%
\makeatletter
\providecommand \@ifxundefined [1]{%
 \@ifx{#1\undefined}
}%
\providecommand \@ifnum [1]{%
 \ifnum #1\expandafter \@firstoftwo
 \else \expandafter \@secondoftwo
 \fi
}%
\providecommand \@ifx [1]{%
 \ifx #1\expandafter \@firstoftwo
 \else \expandafter \@secondoftwo
 \fi
}%
\providecommand \natexlab [1]{#1}%
\providecommand \enquote  [1]{``#1''}%
\providecommand \bibnamefont  [1]{#1}%
\providecommand \bibfnamefont [1]{#1}%
\providecommand \citenamefont [1]{#1}%
\providecommand \href@noop [0]{\@secondoftwo}%
\providecommand \href [0]{\begingroup \@sanitize@url \@href}%
\providecommand \@href[1]{\@@startlink{#1}\@@href}%
\providecommand \@@href[1]{\endgroup#1\@@endlink}%
\providecommand \@sanitize@url [0]{\catcode `\\12\catcode `\$12\catcode
  `\&12\catcode `\#12\catcode `\^12\catcode `\_12\catcode `\%12\relax}%
\providecommand \@@startlink[1]{}%
\providecommand \@@endlink[0]{}%
\providecommand \url  [0]{\begingroup\@sanitize@url \@url }%
\providecommand \@url [1]{\endgroup\@href {#1}{\urlprefix }}%
\providecommand \urlprefix  [0]{URL }%
\providecommand \Eprint [0]{\href }%
\providecommand \doibase [0]{http://dx.doi.org/}%
\providecommand \selectlanguage [0]{\@gobble}%
\providecommand \bibinfo  [0]{\@secondoftwo}%
\providecommand \bibfield  [0]{\@secondoftwo}%
\providecommand \translation [1]{[#1]}%
\providecommand \BibitemOpen [0]{}%
\providecommand \bibitemStop [0]{}%
\providecommand \bibitemNoStop [0]{.\EOS\space}%
\providecommand \EOS [0]{\spacefactor3000\relax}%
\providecommand \BibitemShut  [1]{\csname bibitem#1\endcsname}%
\let\auto@bib@innerbib\@empty
\bibitem [{\citenamefont {Moore}(1998)}]{658762}%
  \BibitemOpen
  \bibfield  {author} {\bibinfo {author} {\bibfnamefont {G.~E.}\ \bibnamefont
  {Moore}},\ }\href {\doibase 10.1109/JPROC.1998.658762} {\bibfield  {journal}
  {\bibinfo  {journal} {Proceedings of the IEEE}\ }\textbf {\bibinfo {volume}
  {86}},\ \bibinfo {pages} {82} (\bibinfo {year} {1998})}\BibitemShut {NoStop}%
\bibitem [{\citenamefont {Kim}\ \emph {et~al.}(2003)\citenamefont {Kim},
  \citenamefont {Austin}, \citenamefont {Baauw}, \citenamefont {Mudge},
  \citenamefont {Flautner}, \citenamefont {Hu}, \citenamefont {Irwin},
  \citenamefont {Kandemir},\ and\ \citenamefont {Narayanan}}]{1250885}%
  \BibitemOpen
  \bibfield  {author} {\bibinfo {author} {\bibfnamefont {N.~S.}\ \bibnamefont
  {Kim}}, \bibinfo {author} {\bibfnamefont {T.}~\bibnamefont {Austin}},
  \bibinfo {author} {\bibfnamefont {D.}~\bibnamefont {Baauw}}, \bibinfo
  {author} {\bibfnamefont {T.}~\bibnamefont {Mudge}}, \bibinfo {author}
  {\bibfnamefont {K.}~\bibnamefont {Flautner}}, \bibinfo {author}
  {\bibfnamefont {J.~S.}\ \bibnamefont {Hu}}, \bibinfo {author} {\bibfnamefont
  {M.~J.}\ \bibnamefont {Irwin}}, \bibinfo {author} {\bibfnamefont
  {M.}~\bibnamefont {Kandemir}}, \ and\ \bibinfo {author} {\bibfnamefont
  {V.}~\bibnamefont {Narayanan}},\ }\href {\doibase 10.1109/MC.2003.1250885}
  {\bibfield  {journal} {\bibinfo  {journal} {Computer}\ }\textbf {\bibinfo
  {volume} {36}},\ \bibinfo {pages} {68} (\bibinfo {year} {2003})}\BibitemShut
  {NoStop}%
\bibitem [{\citenamefont {Salahuddin}\ and\ \citenamefont
  {Datta}(2008)}]{doi:10.1021/nl071804g}%
  \BibitemOpen
  \bibfield  {author} {\bibinfo {author} {\bibfnamefont {S.}~\bibnamefont
  {Salahuddin}}\ and\ \bibinfo {author} {\bibfnamefont {S.}~\bibnamefont
  {Datta}},\ }\href {\doibase 10.1021/nl071804g} {\bibfield  {journal}
  {\bibinfo  {journal} {Nano Letters}\ }\textbf {\bibinfo {volume} {8}},\
  \bibinfo {pages} {405} (\bibinfo {year} {2008})},\ \bibinfo {note} {pMID:
  18052402},\ \Eprint
  {http://arxiv.org/abs/http://dx.doi.org/10.1021/nl071804g}
  {http://dx.doi.org/10.1021/nl071804g} \BibitemShut {NoStop}%
\bibitem [{\citenamefont {Chang}\ \emph
  {et~al.}(2017{\natexlab{a}})\citenamefont {Chang}, \citenamefont {Avci},
  \citenamefont {Nikonov},\ and\ \citenamefont {Young}}]{8027211}%
  \BibitemOpen
  \bibfield  {author} {\bibinfo {author} {\bibfnamefont {S.~C.}\ \bibnamefont
  {Chang}}, \bibinfo {author} {\bibfnamefont {U.~E.}\ \bibnamefont {Avci}},
  \bibinfo {author} {\bibfnamefont {D.~E.}\ \bibnamefont {Nikonov}}, \ and\
  \bibinfo {author} {\bibfnamefont {I.~A.}\ \bibnamefont {Young}},\ }\href
  {\doibase 10.1109/JXCDC.2017.2750108} {\bibfield  {journal} {\bibinfo
  {journal} {IEEE Journal on Exploratory Solid-State Computational Devices and
  Circuits}\ }\textbf {\bibinfo {volume} {PP}},\ \bibinfo {pages} {1} (\bibinfo
  {year} {2017}{\natexlab{a}})}\BibitemShut {NoStop}%
\bibitem [{\citenamefont {Khan}\ \emph {et~al.}(2015)\citenamefont {Khan},
  \citenamefont {Chatterjee}, \citenamefont {Wang}, \citenamefont {Drapcho},
  \citenamefont {You}, \citenamefont {Serrao}, \citenamefont {Bakaul},
  \citenamefont {Ramesh},\ and\ \citenamefont {Salahuddin}}]{Khan2015}%
  \BibitemOpen
  \bibfield  {author} {\bibinfo {author} {\bibfnamefont {A.~I.}\ \bibnamefont
  {Khan}}, \bibinfo {author} {\bibfnamefont {K.}~\bibnamefont {Chatterjee}},
  \bibinfo {author} {\bibfnamefont {B.}~\bibnamefont {Wang}}, \bibinfo {author}
  {\bibfnamefont {S.}~\bibnamefont {Drapcho}}, \bibinfo {author} {\bibfnamefont
  {L.}~\bibnamefont {You}}, \bibinfo {author} {\bibfnamefont {C.}~\bibnamefont
  {Serrao}}, \bibinfo {author} {\bibfnamefont {S.~R.}\ \bibnamefont {Bakaul}},
  \bibinfo {author} {\bibfnamefont {R.}~\bibnamefont {Ramesh}}, \ and\ \bibinfo
  {author} {\bibfnamefont {S.}~\bibnamefont {Salahuddin}},\ }\href
  {http://dx.doi.org/10.1038/nmat4148} {\bibfield  {journal} {\bibinfo
  {journal} {Nat Mater}\ }\textbf {\bibinfo {volume} {14}},\ \bibinfo {pages}
  {182} (\bibinfo {year} {2015})},\ \bibinfo {note} {letter}\BibitemShut
  {NoStop}%
\bibitem [{\citenamefont {Hoffmann}\ \emph {et~al.}(2016)\citenamefont
  {Hoffmann}, \citenamefont {Pešić}, \citenamefont {Chatterjee},
  \citenamefont {Khan}, \citenamefont {Salahuddin}, \citenamefont {Slesazeck},
  \citenamefont {Schroeder},\ and\ \citenamefont
  {Mikolajick}}]{ADFM:ADFM201602869}%
  \BibitemOpen
  \bibfield  {author} {\bibinfo {author} {\bibfnamefont {M.}~\bibnamefont
  {Hoffmann}}, \bibinfo {author} {\bibfnamefont {M.}~\bibnamefont {Pešić}},
  \bibinfo {author} {\bibfnamefont {K.}~\bibnamefont {Chatterjee}}, \bibinfo
  {author} {\bibfnamefont {A.~I.}\ \bibnamefont {Khan}}, \bibinfo {author}
  {\bibfnamefont {S.}~\bibnamefont {Salahuddin}}, \bibinfo {author}
  {\bibfnamefont {S.}~\bibnamefont {Slesazeck}}, \bibinfo {author}
  {\bibfnamefont {U.}~\bibnamefont {Schroeder}}, \ and\ \bibinfo {author}
  {\bibfnamefont {T.}~\bibnamefont {Mikolajick}},\ }\href {\doibase
  10.1002/adfm.201602869} {\bibfield  {journal} {\bibinfo  {journal} {Advanced
  Functional Materials}\ }\textbf {\bibinfo {volume} {26}},\ \bibinfo {pages}
  {8643} (\bibinfo {year} {2016})}\BibitemShut {NoStop}%
\bibitem [{\citenamefont {Kretschmer}\ and\ \citenamefont
  {Binder}(1979)}]{PhysRevB.20.1065}%
  \BibitemOpen
  \bibfield  {author} {\bibinfo {author} {\bibfnamefont {R.}~\bibnamefont
  {Kretschmer}}\ and\ \bibinfo {author} {\bibfnamefont {K.}~\bibnamefont
  {Binder}},\ }\href {\doibase 10.1103/PhysRevB.20.1065} {\bibfield  {journal}
  {\bibinfo  {journal} {Phys. Rev. B}\ }\textbf {\bibinfo {volume} {20}},\
  \bibinfo {pages} {1065} (\bibinfo {year} {1979})}\BibitemShut {NoStop}%
\bibitem [{\citenamefont {Chang}\ \emph {et~al.}(2016)\citenamefont {Chang},
  \citenamefont {Manipatruni}, \citenamefont {Nikonov},\ and\ \citenamefont
  {Young}}]{7373582}%
  \BibitemOpen
  \bibfield  {author} {\bibinfo {author} {\bibfnamefont {S.~C.}\ \bibnamefont
  {Chang}}, \bibinfo {author} {\bibfnamefont {S.}~\bibnamefont {Manipatruni}},
  \bibinfo {author} {\bibfnamefont {D.~E.}\ \bibnamefont {Nikonov}}, \ and\
  \bibinfo {author} {\bibfnamefont {I.~A.}\ \bibnamefont {Young}},\ }\href
  {\doibase 10.1109/JXCDC.2016.2515120} {\bibfield  {journal} {\bibinfo
  {journal} {IEEE Journal on Exploratory Solid-State Computational Devices and
  Circuits}\ }\textbf {\bibinfo {volume} {2}},\ \bibinfo {pages} {1} (\bibinfo
  {year} {2016})}\BibitemShut {NoStop}%
\bibitem [{\citenamefont {Qi}\ and\ \citenamefont
  {Rappe}(2015)}]{PhysRevApplied.4.044014}%
  \BibitemOpen
  \bibfield  {author} {\bibinfo {author} {\bibfnamefont {Y.}~\bibnamefont
  {Qi}}\ and\ \bibinfo {author} {\bibfnamefont {A.~M.}\ \bibnamefont {Rappe}},\
  }\href {\doibase 10.1103/PhysRevApplied.4.044014} {\bibfield  {journal}
  {\bibinfo  {journal} {Phys. Rev. Applied}\ }\textbf {\bibinfo {volume} {4}},\
  \bibinfo {pages} {044014} (\bibinfo {year} {2015})}\BibitemShut {NoStop}%
\bibitem [{\citenamefont {Liu}\ \emph {et~al.}(2013)\citenamefont {Liu},
  \citenamefont {Lou}, \citenamefont {Bibes},\ and\ \citenamefont
  {Dkhil}}]{PhysRevB.88.024106}%
  \BibitemOpen
  \bibfield  {author} {\bibinfo {author} {\bibfnamefont {Y.}~\bibnamefont
  {Liu}}, \bibinfo {author} {\bibfnamefont {X.}~\bibnamefont {Lou}}, \bibinfo
  {author} {\bibfnamefont {M.}~\bibnamefont {Bibes}}, \ and\ \bibinfo {author}
  {\bibfnamefont {B.}~\bibnamefont {Dkhil}},\ }\href {\doibase
  10.1103/PhysRevB.88.024106} {\bibfield  {journal} {\bibinfo  {journal} {Phys.
  Rev. B}\ }\textbf {\bibinfo {volume} {88}},\ \bibinfo {pages} {024106}
  (\bibinfo {year} {2013})}\BibitemShut {NoStop}%
\bibitem [{\citenamefont {Vizdrik}\ \emph {et~al.}(2003)\citenamefont
  {Vizdrik}, \citenamefont {Ducharme}, \citenamefont {Fridkin},\ and\
  \citenamefont {Yudin}}]{PhysRevB.68.094113}%
  \BibitemOpen
  \bibfield  {author} {\bibinfo {author} {\bibfnamefont {G.}~\bibnamefont
  {Vizdrik}}, \bibinfo {author} {\bibfnamefont {S.}~\bibnamefont {Ducharme}},
  \bibinfo {author} {\bibfnamefont {V.~M.}\ \bibnamefont {Fridkin}}, \ and\
  \bibinfo {author} {\bibfnamefont {S.~G.}\ \bibnamefont {Yudin}},\ }\href
  {\doibase 10.1103/PhysRevB.68.094113} {\bibfield  {journal} {\bibinfo
  {journal} {Phys. Rev. B}\ }\textbf {\bibinfo {volume} {68}},\ \bibinfo
  {pages} {094113} (\bibinfo {year} {2003})}\BibitemShut {NoStop}%
\bibitem [{\citenamefont {Chang}\ \emph
  {et~al.}(2017{\natexlab{b}})\citenamefont {Chang}, \citenamefont {Naeemi},
  \citenamefont {Nikonov},\ and\ \citenamefont
  {Gruverman}}]{PhysRevApplied.7.024005}%
  \BibitemOpen
  \bibfield  {author} {\bibinfo {author} {\bibfnamefont {S.-C.}\ \bibnamefont
  {Chang}}, \bibinfo {author} {\bibfnamefont {A.}~\bibnamefont {Naeemi}},
  \bibinfo {author} {\bibfnamefont {D.~E.}\ \bibnamefont {Nikonov}}, \ and\
  \bibinfo {author} {\bibfnamefont {A.}~\bibnamefont {Gruverman}},\ }\href
  {\doibase 10.1103/PhysRevApplied.7.024005} {\bibfield  {journal} {\bibinfo
  {journal} {Phys. Rev. Applied}\ }\textbf {\bibinfo {volume} {7}},\ \bibinfo
  {pages} {024005} (\bibinfo {year} {2017}{\natexlab{b}})}\BibitemShut
  {NoStop}%
\bibitem [{\citenamefont {Aziz}\ \emph {et~al.}(2016)\citenamefont {Aziz},
  \citenamefont {Ghosh}, \citenamefont {Datta},\ and\ \citenamefont
  {Gupta}}]{7458805}%
  \BibitemOpen
  \bibfield  {author} {\bibinfo {author} {\bibfnamefont {A.}~\bibnamefont
  {Aziz}}, \bibinfo {author} {\bibfnamefont {S.}~\bibnamefont {Ghosh}},
  \bibinfo {author} {\bibfnamefont {S.}~\bibnamefont {Datta}}, \ and\ \bibinfo
  {author} {\bibfnamefont {S.~K.}\ \bibnamefont {Gupta}},\ }\href {\doibase
  10.1109/LED.2016.2558149} {\bibfield  {journal} {\bibinfo  {journal} {IEEE
  Electron Device Letters}\ }\textbf {\bibinfo {volume} {37}},\ \bibinfo
  {pages} {805} (\bibinfo {year} {2016})}\BibitemShut {NoStop}%
\bibitem [{\citenamefont {Catalan}\ \emph {et~al.}(2015)\citenamefont
  {Catalan}, \citenamefont {Jimenez},\ and\ \citenamefont
  {Gruverman}}]{Catalan2015}%
  \BibitemOpen
  \bibfield  {author} {\bibinfo {author} {\bibfnamefont {G.}~\bibnamefont
  {Catalan}}, \bibinfo {author} {\bibfnamefont {D.}~\bibnamefont {Jimenez}}, \
  and\ \bibinfo {author} {\bibfnamefont {A.}~\bibnamefont {Gruverman}},\ }\href
  {http://dx.doi.org/10.1038/nmat4195} {\bibfield  {journal} {\bibinfo
  {journal} {Nat Mater}\ }\textbf {\bibinfo {volume} {14}},\ \bibinfo {pages}
  {137} (\bibinfo {year} {2015})},\ \bibinfo {note} {news and
  Views}\BibitemShut {NoStop}%
\bibitem [{\citenamefont {Stamm}\ \emph {et~al.}(2013)\citenamefont {Stamm},
  \citenamefont {Kim}, \citenamefont {Lu}, \citenamefont {Bark}, \citenamefont
  {Eom},\ and\ \citenamefont {Gruverman}}]{doi:10.1063/1.4794865}%
  \BibitemOpen
  \bibfield  {author} {\bibinfo {author} {\bibfnamefont {A.}~\bibnamefont
  {Stamm}}, \bibinfo {author} {\bibfnamefont {D.~J.}\ \bibnamefont {Kim}},
  \bibinfo {author} {\bibfnamefont {H.}~\bibnamefont {Lu}}, \bibinfo {author}
  {\bibfnamefont {C.~W.}\ \bibnamefont {Bark}}, \bibinfo {author}
  {\bibfnamefont {C.~B.}\ \bibnamefont {Eom}}, \ and\ \bibinfo {author}
  {\bibfnamefont {A.}~\bibnamefont {Gruverman}},\ }\href {\doibase
  10.1063/1.4794865} {\bibfield  {journal} {\bibinfo  {journal} {Applied
  Physics Letters}\ }\textbf {\bibinfo {volume} {102}},\ \bibinfo {pages}
  {092901} (\bibinfo {year} {2013})},\ \Eprint
  {http://arxiv.org/abs/http://dx.doi.org/10.1063/1.4794865}
  {http://dx.doi.org/10.1063/1.4794865} \BibitemShut {NoStop}%
\end{thebibliography}%
	
\end{document}